# Stoichiometry control and epitaxial growth of AgCrSe$_2$ thin films by pulsed-laser deposition

Yusuke Tajima[1,*], Kenshin Inamura[1,*], Sebun Masaki[1], Takumi Yamazaki[2],

Takeshi Seki[2,3], Kazutaka Kudo[1,4], Jobu Matsuno[1,4], and Junichi Shiogai[1,4,†]

[1]*Department of Physics, Osaka University, Toyonaka, Osaka 560-0043, Japan*

[2]*Institute for Materials Research, Tohoku University, Sendai, Miyagi 980-8577, Japan*

[3]*Center for Science and Innovation in Spintronics, Tohoku University, Sendai, Miyagi 980-8577, Japan*

[4]*Division of Spintronics Research Network, Institute for Open and Transdisciplinary Research Initiatives, Osaka University, Suita, Osaka 565-0871, Japan*

[†]Electronic mail: junichi.shiogai.sci@osaka-u.ac.jp

[*]These authors equally contribute to this work

Abstract

We report on epitaxial growth in thin-film synthesis of a polar magnetic semiconductor $AgCrSe_2$ on lattice-matched yttria-stabilized zirconia (111) substrate by pulsed-layer deposition (PLD). By using Ag-rich PLD target to compensate for Ag deficiency in thin films, the nucleation of impurity phases is suppressed, resulting in the *c*-axis-oriented and single-phase $AgCrSe_2$ thin film. Structural analysis using x-ray diffraction and cross-sectional scanning transmission electron microscopy reveals epitaxial growth with the presence of both twisted and polar domains. Optical absorbance spectrum and magnetization measurements show absorption edge at around 0.84 eV and magnetic transition temperature at 41 K, respectively. These values are consistent with the reported values of direct bandgap and Néel temperature of bulk $AgCrSe_2$, reflecting a single-phase and stoichiometric feature of the obtained film. Our demonstration of epitaxial thin-film growth of $AgCrSe_2$ serves as a bedrock for exploration of its potential thermoelectric and spintronic functionalities at surface or heterointerfaces.

I. INTRODUCTION

A ternary chromium compound AgCrSe$_2$ ($R3m$) is a $p$-type and polar magnetic semiconductor [1]. Its hexagonal crystal structure with $a$ and $c$-axis length being 3.680 and 21.225 Å is composed of alternating stacking of two-dimensional (2D) network of edge-sharing CrSe$_6$ octahedra (referred as CrSe$_2$ network, hereafter) and AgSe$_4$ tetrahedral layer [2][3] as shown in the left panel of Fig. 1(a). AgCrSe$_2$ and related family compounds $A$Cr$X_2$ ($A$: monovalent ions, $X$: chalcogens) have recently attracted much interest in a wide range of realms because they exhibit unique structural and ionic properties related to $A$ sublattices [4][5], and intriguing magnetic state [6][7] and magneto-transport phenomena [8][9] stemming from the Cr triangular lattice in the Cr$X_2$ network. Previous powder x-ray diffraction (XRD) study has revealed that AgCrSe$_2$ undergoes a second-order transition from low-temperature ordered phase ($R3m$) to high-temperature disordered phase [$R\bar{3}m$, the right panel of Fig. 1(a)] across a temperature around 475 K [3]. This transition is related to disordering of Ag occupancy at the $\alpha$ and $\beta$ tetrahedral sites between the CrSe$_2$ networks [Fig. 1(b)]. In the low-temperature phase, the Ag occupies either $\alpha$ or $\beta$ site and forms non-buckled one-atomic-layer 2D sheet. The tetrahedral coordination of Ag breaks inversion symmetry along the $c$ axis and induces spontaneous polarization: three bonds are formed with upper Se atoms and one bond is formed with lower Se atom when Ag occupies the $\alpha$ site as shown in Fig. 1(b). The polarization is inverted when Ag occupies the $\beta$ site. The spontaneous polarization gives rise to the Rashba spin-orbit interaction, which induces spin-split metallic surface state on the $CrSe_2$ termination [10] or giant magnetoresistance in bulk [8]. In addition, $S = 3/2$ spin of $Cr^{3+}$ within the $CrSe_2$ network is regarded as 2D triangular-lattice antiferromagnet with Néel temperature $T_N$ below 50 K in $AgCrSe_2$, leading to exotic magnetic states such as cycloidal [2][11] or noncoplanar spin structures [6][12]. Actually, the Weiss temperature of around 75 K being higher than $T_N$ is reported in $AgCrSe_2$ as a signature of spin frustration [6]. In the high-temperature phase, Ag occupies 50% of each site and forms 2D buckled honeycomb lattice in the *ab* plane. The superionic behavior [1][13] and ultralow thermal conductivity [4], linked to the disordering of Ag ions [5][14], has been reported. On this basis, $AgCrSe_2$ is also a promising thermoelectric material [15][16].

Considering that intensive studies of $AgCrSe_2$ in bulk and powder forms have revealed its high potential not only in fundamental spin-related physical phenomena but also for energy-harvesting application, thin-film synthesis of $AgCrSe_2$ will make this family compounds even more attractive in views of large-area application, observation of quantum phenomena by electric-field gating [17] or amplifying surface effect [10], as well as possible emergence of multiferroicity in ultra-thin limit [18]. Yet, synthesis of single-phase AgCrSe$_2$ thin film remains to be accomplished probably because of difficulty in composition tuning owing to highly volatile feature of Ag and tendency for segregation of Cr-Se impurity phases [19]. In this study, we report on synthesis of epitaxial and single-phase AgCrSe$_2$ thin film in pulsed-laser deposition (PLD) using Ag-rich target, which we found allows compensating for Ag deficiency and suppressing the impurity phases.

Our previous work suggests that the (111) plane of yttria-stabilized zirconia (YSZ: $Fm\bar{3}m$) substrate is suitable for epitaxial thin-film growth of the CrSe$_2$ network [20]. On the CrSe$_2$ network of AgCrSe$_2$, the Se atoms of the CrSe$_6$ octahedra form a triangular lattice as shown in Fig. 1(c) and the neighboring Se-Se distance is 3.680 Å [3]. Figure 1(d) illustrates in-plane atomic arrangement of the top-most O atoms and underlying cations (Y or Zr) on YSZ(111) and considers possible epitaxial relationships, which are attributed to twisted and polar domains of AgCrSe$_2$ ($R3m$). The neighboring O-O distance of the top-most O plane on YSZ(111) is 3.639 Å, which closely matches with the Se-Se distance with a small lattice mismatch of 1.1%. In the case that the $c$ axis of AgCrSe$_2$ points upwards (+$Z$ polar, orange rhombus), the in-plane orientation of the Se triangular lattice consists of two possible configurations: the $a$ axis of the CrSe$_2$ network [red arrow in Fig. 1(d)] is parallel to the [1$\bar{1}$0] direction of YSZ(111) (denoted as A domain) and is rotated by 60° with the *a* axis being parallel to $[10\bar{1}]$ of YSZ(111) (denoted as B domain). Despite the three-fold symmetry around the *c* axis of $AgCrSe_2$ and [111] of YSZ(111), the twisted domains in thin film seem energetically degenerate owing to the structural similarity of the Se and O triangular lattices. The same argument was also given for the thin-film growth of the similar compound, delafossite oxide $PdCoO_2$ on $Al_2O_3$(001) substrate [21]. By comparing with $PdCoO_2$ having O-Pd-O dumbbell-like bonding between the $CoO_2$ networks, $AgCrSe_2$ possesses $AgSe_4$ tetrahedral layer between the $CrSe_2$ networks, which breaks inversion symmetry. Thus, extra two configurations, which rely on this tetrahedral coordination, should be considered for the low-temperature ordered phase: the *c* axis of $AgCrSe_2$ points upwards (denoted as +*Z* polar) or downwards (denoted as −*Z* polar, blue rhombus) as shown in Fig. 1(d). In the structural characterization of $AgCrSe_2$ thin film, these two degrees of freedom, in-plane rotation and polarization, are considered.

## II. EXPERIMENTAL METHODS

The 100-nm-thick thin-film samples were prepared on YSZ(111) substrates by pulsed-laser deposition at substrate temperature $T_{sub}$ ranging from 300 to 500°C. We used an Ag-rich $AgCrSe_2$ target for synthesizing stoichiometric thin films and a stoichiometric

AgCrSe$_2$ target as a reference experiment. For synthesis of the PLD targets, we first prepared polycrystalline AgCrSe$_2$ samples using a solid-state reaction. Stoichiometric amounts of the starting materials Ag (99.9%), Cr (99.99%), and Se (99.9%) (Kojundo Chemical Laboratory Co. Ltd) were mixed, pressed into pellets, and then heated in an evacuated quartz tube at 800ºC for 24h twice with pulverizing and pressing in between. The resulting samples were characterized by means of powder X-ray diffraction (XRD) using an X-ray diffractometer (Rigaku MiniFlex600-C) with Cu-$K\alpha$ radiation, equipped with a high-speed one-dimensional detector (Rigaku D/teX Ultra2) and it was confirmed that they consisted of single-phase AgCrSe$_2$. The obtained polycrystalline AgCrSe$_2$ pellet was then used as the stoichiometric target. The Ag-rich target was prepared by grinding the stoichiometric AgCrSe$_2$ samples, mixing them with commercially-available Ag$_2$Se ($P2_12_12_1$) powder (Kojundo Chemical Laboratory Co. Ltd) with a 2:1 molar ratio, and pressing into a pellet. The resultant target consists of a mixture of AgCrSe$_2$ and Ag$_2$Se with the Ag/Cr atomic ratio of 2. The structural characterization of thin-film samples was performed by XRD (X'pert PRO MRD, PANalytical B.V.) with Cu-$K\alpha_1$ X-ray source ($\lambda$ = 1.5406 Å) and cross-sectional scanning transmission electron microscopy (STEM) imaging using FEI-Titan G2-cubed (TU-503). The Ag/Cr chemical composition ratio of thin films was determined by energy dispersive X-ray spectroscopy (EDX) using a scanning electron microscope (Hitachi High-Tech TM4000Plus II) equipped with an energy dispersive spectrometer (Oxford Instruments AztecOne). The optical absorption spectrum of the AgCrSe$_2$ thin film was obtained by the total reflectance measurement using an ultraviolet-visible spectrophotometer equipped with integrating sphere (U-4100, HITACHI, Japan). Magnetization measurements were performed by magnetic properties measurement system (MPMS3, Quantum Design, Inc.) with a typical sample area of about 3×3 mm$^2$. Schematics of crystal structure, in-plane atomic arrangements, and cross-sectional structural patterns were drawn with VESTA3 [22].

## III. RESULTS AND DISCUSSION

**A. Optimization of growth condition for single-crystalline AgCrSe$_2$ thin film**

We first optimize growth condition for synthesis of single-phase AgCrSe$_2$ thin film. Figure 2(a) shows a 2theta-omega scan of the XRD patterns of the AgCrSe$_2$ thin films grown at $T_{sub}$ = 450 and 500 °C using the stoichiometric AgCrSe$_2$ target (black lines) and those grown at $T_{sub}$ = 400, 425, 450, 475, and 500°C using the Ag-rich target. The main diffraction peaks commonly observed for all the films are identified as diffraction from AgCrSe$_2$(003$l$), indicating the *c*-axis-oriented growth of $R3m$ structure since the low-angle (003) and (006) diffraction peaks are prohibited for the $R\bar{3}m$ structure [4][5]. For the films using stoichiometric target, the larger intensity of the (003) and (00$\underline{12}$) diffraction peaks of $Cr_2Se_3$ than (00$\underline{3}l$) of $AgCrSe_2$ are observed, indicating the presence of significant volume fraction of the $Cr_2Se_3$ phase. Such segregation of the $Cr_2Se_3$ phase was reported in the previous thin-film study using molecular beam epitaxy and the emergence of this Ag-deficient impurity phase was attributed to the high volatility of Ag [19]. The Ag-deficient phase is common impurity in $AgCrSe_2$ in both bulk and thin films [19,23]. In addition, a weak Ag(331) peak was also observed around 2theta ~ 78°. This may be caused by the aggregation of Ag expelled from the Ag-deficient phase. With the aim of suppression of the $Cr_2Se_3$ phase, we employed the Ag-rich target to compensate for the Ag deficiency. At $T_{sub}$ = 400°C [purple line in Fig. 2(a)], the weak diffraction peaks of $AgCrSe_2$ are observed while those of $Cr_2Se_3$ are not detected. With increasing $T_{sub}$ up to 450 [red line in Fig. 2(a)], the intensity of the $AgCrSe_2$ peaks becomes stronger than those for 400 and 425°C with keeping the absence of the impurity phases. With further increasing $T_{sub}$ to 475 and 500°C [light and dark orange lines in Fig. 2(a)], the $Cr_2Se_3$ peaks start to appear. The *c*-axis length determined by the (009) peak is 21.226 ± 0.005 Å, which shows no systematic variation among samples prepared with different $T_{sub}$ and is almost consistent with the bulk value (*c* = 21.225 Å) [3]. Figure 2(b) shows the $T_{sub}$ dependence of Ag/Cr composition ratio for the thin films fabricated from the stoichiometric and Ag-rich PLD targets. In the whole $T_{sub}$ range, the Ag/Cr ratio is smaller for the stoichiometric target than that for the Ag-rich target. In addition, with increasing $T_{sub}$, the Ag/Cr ratio for the Ag-rich target starts to decrease in $T_{sub} \geq 475°C$, which can be explained by the high volatility of Ag and is consistent with the emergence of the $Cr_2Se_3$ phase in thin films grown at high $T_{sub}$ as shown in Fig. 2(a). The stoichiometric composition of $AgCrSe_2$ is obtained in $T_{sub} \leq 450°C$ [a horizontal dashed line in Fig. 2(b)] only when using the Ag-rich target, which is also consistent with the single phase $AgCrSe_2$ in this growth condition as judged by XRD. Based on these observations, we conclude that $T_{sub} = 450°C$ is the optimal growth temperature for obtaining the single-phase $AgCrSe_2$ thin film when using the Ag-rich target with Ag/Cr ratio of 2. The optimal $T_{sub}$ could be made higher by using the target containing a higher fraction of $Ag_2Se$. Hereafter, further structural and physical properties characterizations were performed for the $AgCrSe_2$ thin film grown at $T_{sub} = 450°C$ with the Ag-rich target.

Figure 2(c) shows the in-plane phi-scan XRD pattern of $AgCrSe_2(101)$ and YSZ(400) diffraction peaks. Three-fold in-plane rotational symmetry is exhibited in YSZ(400), being consistent with the $C_3$ symmetry around the [111] axis on YSZ(111). On the other hand, we observe six-fold peaks for $AgCrSe_2(101)$ every 60°, despite that the [001] axis of the $R3m$ structure is the $C_3$ symmetry. The $a$-axis length of 3.682 Å extracted from AgCrSe$_2$(101) is consistent with the bulk value ($a$ = 3.680 Å). From these in-plane XRD pattern, we conclude that the $c$-axis-oriented AgCrSe$_2$ thin film contains two twisted domains rotated by 60° around the $c$ axis with the epitaxial relationship in which the $a$ axis of AgCrSe$_2$ is parallel to YSZ[1$\bar{1}$0] (A domain) and YSZ[10$\bar{1}$] (B domain) as discussed in Fig. 1(d).

**B. Structural characterization using atomically-resolved STEM imaging**

Cross-sectional STEM imaging provides further microscopic insights into formation of the single-phase $R3m$ structure of AgCrSe$_2$ as well as crystal orientation. A viewing orientation, as depicted by an orange arrow in Fig. 1(d), was chosen opposite to the [1$\bar{1}$0] direction of YSZ(111) because the [1$\bar{1}$0] direction should be parallel to either $a$ or $b$ axis of AgCrSe$_2$ so that we can identify the Ag ion location between CrSe$_2$ layers as shown in Fig. 1(b). We first investigate elemental distribution in a wide area of cross-sectional high angle annular dark field (HAADF) STEM image in Fig. 3(a) for the AgCrSe$_2$ thin film on the YSZ substrate. The uniform distributions of Ag, Cr, and Se with no segregation are demonstrated, being consistent with the XRD observation. Figure 3(b) shows a magnified HAADF-STEM image, taken in a few tens μm distance from Fig. 1(a), which consists of a grain boundary as indicated by white arrows. The $c$-axis-oriented layered structure of AgCrSe$_2$ is clearly seen over a large area of the film region. Inset in Fig. 3(a) shows an atomically-resolved HAADF-STEM image around the film/substrate interface (area #1), indicating the epitaxial growth of AgCrSe$_2$ on the lattice-matched YSZ(111) substrate with a sharp interface. Note that the Ag layer is not clearly visible in the vicinity of the interface. This is possibly caused by large displacement and aggregation of Ag occurring after preparation of STEM sample using focused ion beam as reported in Ge/Ag multilayers [24]. Another possibility is the preferential formation of an Ag-deficient CrSe$_2$ initial layer on the YSZ substrate during thin-film growth. Figure 3(c) shows HAADF-STEM image and EDX elemental mapping of area #2 [indicated in Fig. 3(a)], as one of representative patterns, and compares them with the simulated cross-sectional structural patterns. The HAADF-STEM and EDX mapping show the presence of non-buckled Ag monolayer between the CrSe$_2$ networks, being consistent with $R3m$ structure of AgCrSe$_2$. Considering the location of Ag sites, the atomic structure of area #2 is well reproduced by the structural pattern assuming +$Z$ polar and A domain defined in Fig. 1(d). Figures 3(d) and 3(e) compare HAADF-STEM images taken around two areas (#3 and #4) across the grain boundary together with corresponding simulated structural patterns. The atomic arrangements of the CrSe$_2$ networks and Ag layers in between areas #3 and #4 show mirror-symmetric patterns, corresponding to the 60º-twisted domain with the common polar direction. By considering the location of Ag, the HAADF-STEM images are consistent with the structural pattern simulating −Z polar and A domain for area #3 [Fig. 3(d)] and −Z polar and B domain for area #4 [Fig. 3(e)], respectively. The piezo-response force microscopy or polarized optical microscopy will provide further quantitative characterization of the polar domains. Although we found the use of lattice-matched YSZ(111) substrate and Ag-rich target is effective for obtaining single-phase epitaxial AgCrSe$_2$ thin film, further improvement in the synthesis process is needed for single-domain films. For example, using facet-controlled substrate [25] or polar substrates [26] may be useful for lifting in-plane degeneracy.

**C. Optical bandgap and magnetic transition temperature of AgCrSe$_2$ thin film**

The physical properties of the single-phase AgCrSe$_2$ thin film were characterized in view of optical bandgap $E_g$ and Néel temperature $T_N$. Previous study of Hall measurement in bulk experiments reported that AgCrSe$_2$ is a *p*-type semiconductor with bandgap being 0.5 eV [1] while the density-functional theory calculations suggest a direct and indirect bandgap of 0.7 and 0.1 eV, respectively [8]. The $E_g$ value of the AgCrSe$_2$ thin film was estimated from absorbance spectrum [$A(h\nu)$ spectrum with $h\nu$ being photon energy] collected in total reflectance measurement. Figure 4(a) shows $A(h\nu)$ spectra for the AgCrSe$_2$ thin film on the YSZ(111) substrate (red) and the reference YSZ(111) substrate (gray). An extrapolation to the $A(h\nu)$ spectrum of the YSZ substrate around the absorption edge yields $E_g \sim 3$ eV, which is consistent with previous report [27], suggesting the validity of the experimental setup. For the AgCrSe$_2$ thin film, a clear absorption edge in $h\nu > 1$ eV is observed. An extrapolation of $A(h\nu)$ spectrum in the linear region of 1.2 eV $< h\nu < 1.7$ eV yields $E_g$ of 0.84 eV, which is roughly consistent with the previous report [8].

Figure 4(b) shows temperature dependence of magnetization [$M(T)$ curve] of AgCrSe$_2$ thin film on YSZ substrate measured in out-of-plane magnetic field of $\mu_0 H = 1$ T after 7 T field cooling in the range of $5 < T < 300$ K. The overall $M(T)$ consists of a negative component in high-temperature range and a positive component sharply increasing with decreasing $T$ in low-temperature range, which we assigned to diamagnetic and paramagnetic contribution of the YSZ substrate, respectively. In addition to the substrate contributions, a clear hump around $T \sim 41$ K is observed as highlighted by the vertical dashed line in the inset of Fig. 4(b). This hump structure is absent in $M(T)$ curve of the reference substrate. By considering the reported value of $T_N \sim 45$ K for the bulk AgCrSe$_2$ [8], we ascribed the observed hump in $M(T)$ curve of the AgCrSe$_2$ thin film as a development of antiferromagnetic order. The observation of comparable values of $E_g$ and $T_N$ with bulk reflects the single-phase and stoichiometric feature of the obtained AgCrSe$_2$ thin film.

**IV. SUMMARY AND OUTLOOK**

In summary, we have obtained an epitaxial and single-phase AgCrSe$_2$ thin film on lattice-matched YSZ(111) substrate by employing Ag-rich target and precisely tuning growth temperature to suppress Ag-deficient impurity phases. The XRD analysis reveals the presence of two in-plane twisted domains rotated by 60º with comparable volume fraction. In addition, the atomic-resolved STEM imaging shows lattice-matched epitaxial growth as well as the presence of polar domains. Optical absorbance spectrum and magnetization measurements suggest that the optical bandgap and Néel temperature of the AgCrSe$_2$ thin film are comparable to those of bulk, being consistent with the single-phase and stoichiometric feature of the AgCrSe$_2$ thin film. Our demonstration of epitaxial thin-film growth of AgCrSe$_2$ tailors a platform to explore its potential functionalities based on thermoelectric, spintronic, or multiferroic thin-film devices.

**ACKNOWLEDGMENTS**

STEM observations and analysis were made with the cooperation of Koji Hayasaka, Makoto Nagasako, and Toyohiko J. Konno of Analytical Research Core for Advanced Materials, Institute for Materials Research, Tohoku University. This work was partly supported by Tohoku University Microstructural Characterization Platform in MEXT Advanced Research Infrastructure for Materials and Nanotechnology in Japan (JPMXP1224TU0205) and performed under the GIMRT Program of the Institute for Materials Research, Tohoku University (Proposal No. 202312-CRKEQ-0033). Total reflectance measurements were performed using research equipment shared in MEXT Project for promoting public utilization of advanced research infrastructure (Program for supporting construction of core facilities), Grant No. JPMXS0441200024. This work was supported by JST, PRESTO Grant No. JPMJPR21A8, and JSPS KAKENHI Grant Nos. JP23H01686, JP23K26379, JP22H01182, JP23K22453, and JP24K21531, and Tanikawa Foundation.

chalcogenides for thermoelectricity, Sci. Technol. Adv. Mater. **13**, 053003 (2012).

17. S. J. Kim, J. Zhu, M. M. Piva, M. Schmidt, D. Fartab, A. P. Mackenzie, M. Baenitz, M. Nicklas, H. Rosner, A. M. Cook, R. González-Hernández, L. Šmejkal, and H. Zhang, Observation of the Anomalous Hall Effect in a Layered Polar Semiconductor, Adv. Sci. **11**, 2307306 (2024).

18. Z. Sun, Y. Su, A. Zhi, Z. Gao, X. Han, K. Wu, L. Bao, Y. Huang, Y. Shi, X. Bai, P. Cheng, L. Chen, K. Wu, X. Tian, C. Wu, and B. Feng, Evidence for multiferroicity in single-layer $CuCrSe_2$, Nature Commun. **15**, 4252 (2024).

19. Y. Nanao, C. Bigi, A. Rajan, G. Vinai, D. Dagur, and P. D. C. King, Epitaxial growth of thin films by molecular beam epitaxy, J. Appl. Phys. **135**, 045303 (2024).

20. Y. Tajima, J. Shiogai, M. Ochi, K. Kudo, and J. Matsuno, Engineered substrates for domain control in CrSe thin-film growth: Single-domain formation on lattice-matched YSZ(111) substrate, Jpn. J. Appl. Phys. In press, 10.35848/1347-4065/addb1b

21. T. Harada, K. Fujiwara, and A. Tsukazaki, Highly conductive $PdCoO_2$ ultrathin films for transparent electrodes, APL Mater. **6**, 046107 (2018).

22. K. Momma and F. Izumi, VESTA 3 for three-dimensional visualization of crystal, volumetric and morphology data, J. Appl. Crystallogr. **44**, 1272 (2011).
20

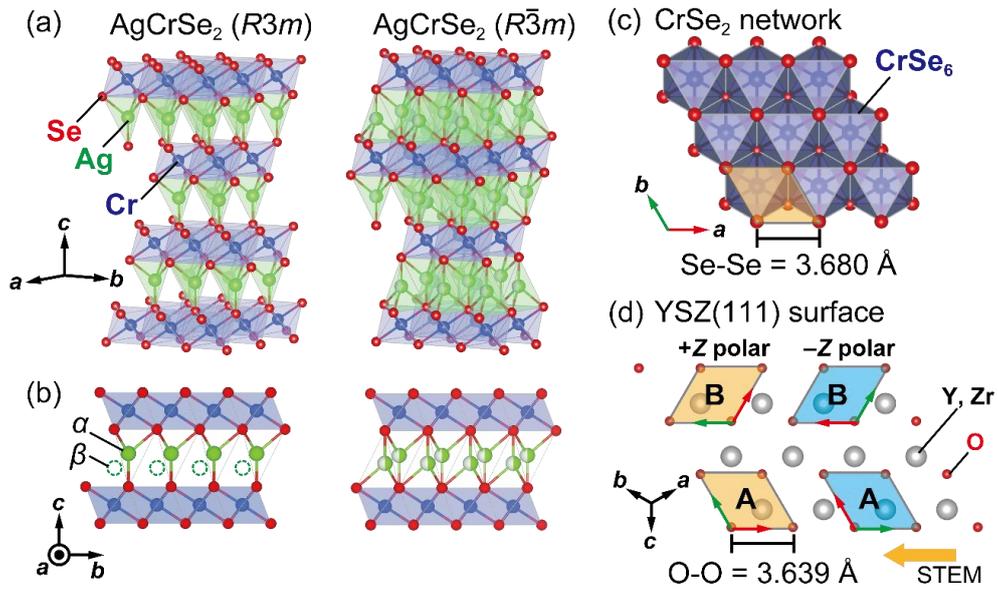

**Figure 1** | (a) Schematic crystal structures of AgCrSe$_2$ in $R3m$ (left) and $R\bar{3}m$ (right) structures. (b) Side views of CrSe$_2$/Ag/CrSe$_2$ stacks in $R3m$ (left) and $R\bar{3}m$ (right) structures. The green and white bicolor spheres represent the half occupancy at Ag sites. (c) Top views of a single layer CrSe$_2$ two-dimensional network. Orange parallelogram represents single unit-cell of the CrSe$_2$ network. (d) In-plane atomic arrangement of the topmost O atoms and Y or Zr atoms on the YSZ(111) surface. Orange parallelograms correspond to the possible arrangement of the CrSe$_2$ network with the epitaxial relationship in which the $a$ axis of AgCrSe$_2$ is parallel to YSZ[$1\bar{1}0$] (denoted as A domain) and YSZ[$10\bar{1}$] (denoted as B domain), and the $c$ axis of AgCrSe$_2$ points +Z direction (+Z polar). Blue parallelograms represent A and B domains with the $c$ axis of AgCrSe$_2$ pointing −Z direction (−Z polar). A horizontal orange arrow indicates observation direction in scanning transmission electron microscopy (STEM) imaging.

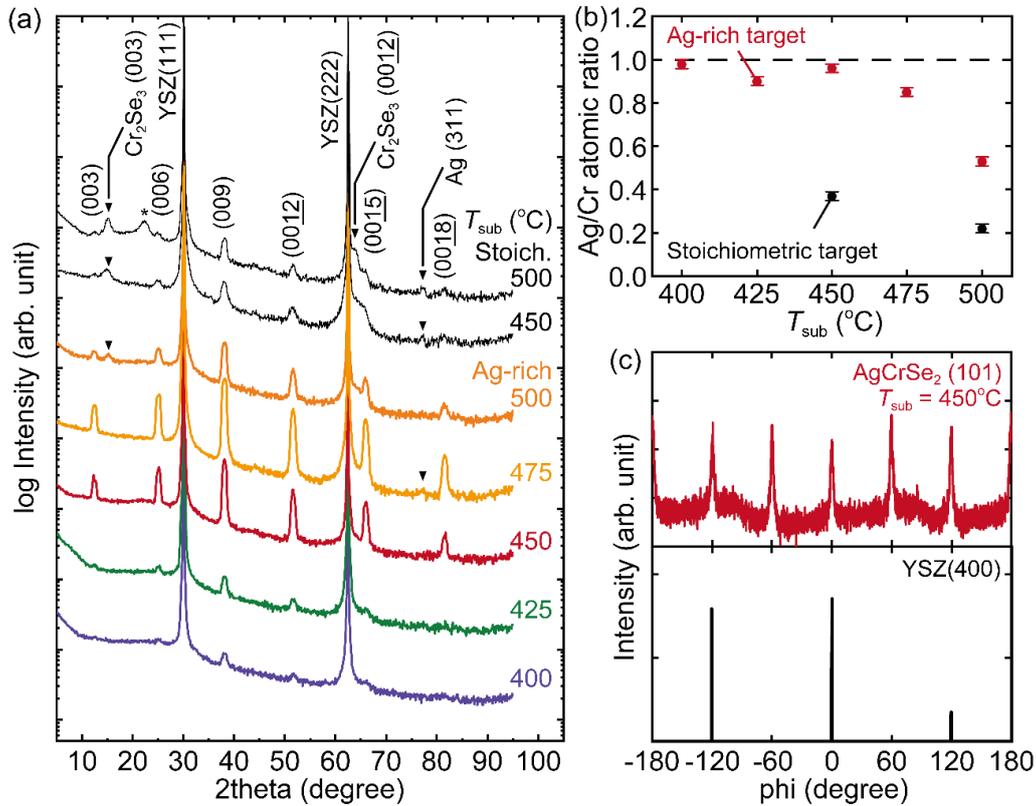

**Figure 2** | (a) 2theta-omega scan of the X-ray diffraction (XRD) patterns for $T_{sub}$ = 450°C and 500°C using stoichiometric $AgCrSe_2$ target (black lines) and those grown at $T_{sub}$ = 400 (purple), 425 (green), 450 (red), 475 (light orange), and 500°C (orange) using Ag-rich target. The asterisk (*) denotes the peak originating from the sample holder. (b) Ag/Cr atomic ratio on the thin films as a function of $T_{sub}$ using stoichiometric (black) and Ag-rich (red) targets. The error bars are estimated from the standard deviations of the Ag/Cr ratio in the different areas of the identical samples. A horizontal dashed line indicate Ag/Cr = 1. (c) In-plane phi-scan of XRD patterns of $AgCrSe_2$(101) (top panel) and YSZ(400) (bottom panel).

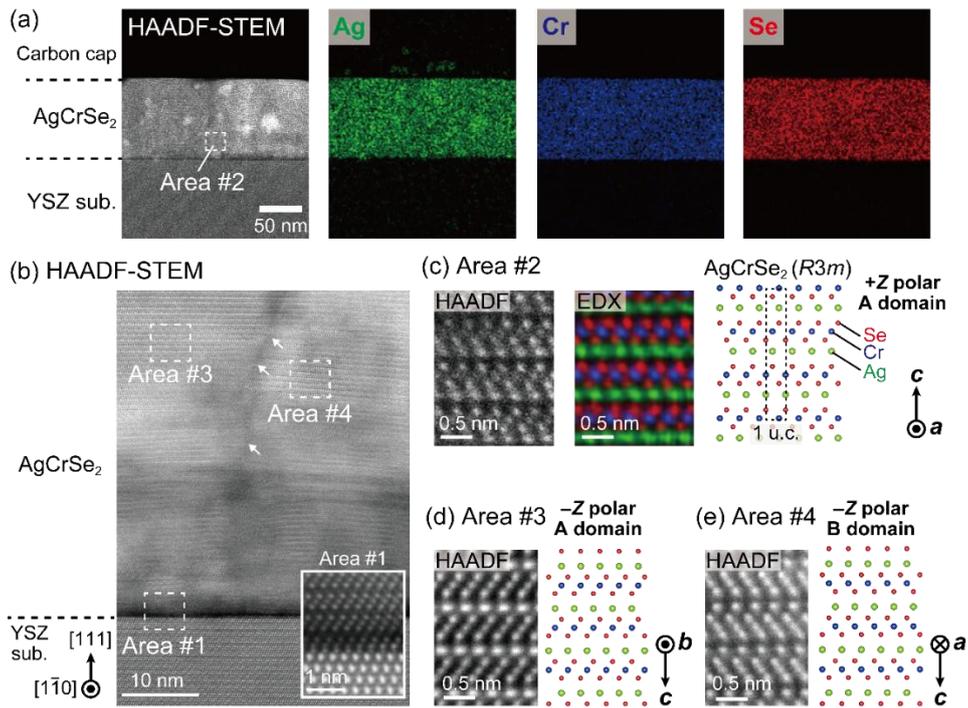

**Figure 3** | (a) Wide-view high-angle annular dark field (HAADF) scanning transmission electron micrograph (STEM) image and elemental distribution mapping of the $AgCrSe_2$ thin film grown on YSZ(111) taken from YSZ[$1\bar{1}0$]. (b) High-resolution HAADF-STEM image of an $AgCrSe_2$ film. White arrows indicate threading dislocation between twisted domains. (Inset) An atomically-resolved HAADF-STEM image around $AgCrSe_2$/YSZ interface (area #1). (c) An atomically-resolved HAADF-STEM image and EDX elemental mapping in area #2, and corresponding cross-sectional structural pattern of the $R3m$ structure viewed from the $a$ axis of $AgCrSe_2$ (+$Z$ polar A domain). (d)(e) Atomically-resolved HAADF-STEM images for areas #3 and #4 and corresponding structural patterns viewed from the $b$ axis of $AgCrSe_2$ (−$Z$ polar A domain) and along the $a$ axis (−$Z$ polar B domain), respectively.

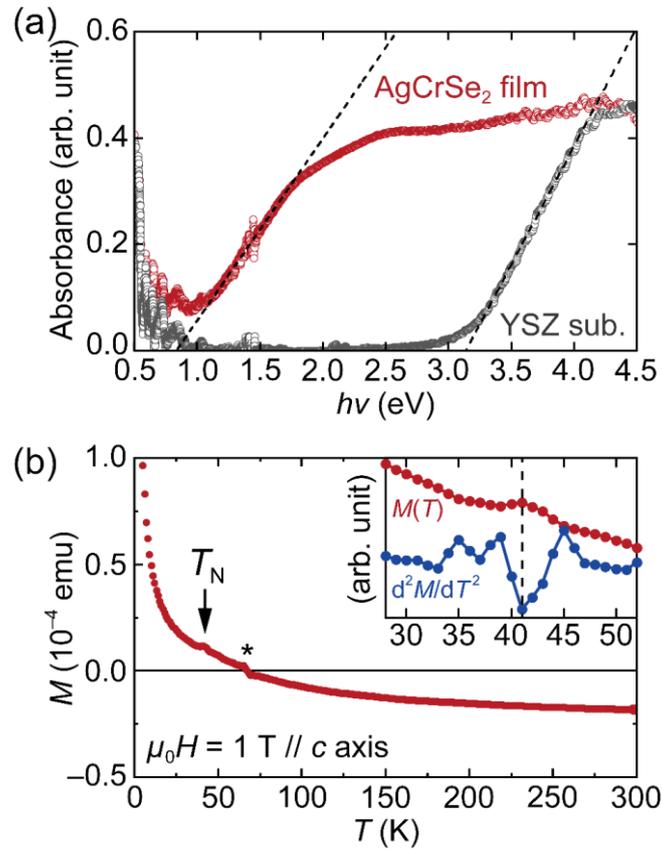

**Figure 4** | (a) Optical absorbance spectra for AgCrSe$_2$ film on YSZ(111) substrate (red) and reference YSZ substrate (gray) obtained from measurement of total reflectance measurement. (b) Temperature dependence of magnetization [$M(T)$ curve] of AgCrSe$_2$ film on YSZ(111) substrate. The $T_N$ and asterisk (*) denote Néel temperature and the temperature where $M(T)$ crosses zero, respectively. (Inset) $M(T)$ and its second derivative around $T_N$.